\documentclass[fleqn,twoside]{article}

\usepackage{espcrc2,url}
\usepackage{graphicx}

\def\s{A0535+26}
\def\mcc#1{\multicolumn{1}{c}{#1}}
\def\Iave{\mbox{$\langle I \rangle$}}
\def\Lx{\mbox{L$_{\rm x}$}}
\newlength{\apex}   

\title{X--ray/optical observations of \s/HDE~245770 in quiescence}

\author{M.~Orlandini\address[IASFBO]{Istituto di Astrofisica Spaziale e
	Fisica Cosmica (IASF/CNR), Sezione Bologna, \\
        \hspace{\apex}Via Gobetti 101, 40129 Bologna, Italy},
        C.~Bartolini\address[INAFBO]{Dipartimento Astronomia,
	Universit\`a di Bologna, \\
	\hspace{\apex}Via Ranzani 1, 40127 Bologna, Italy},
	S.~Campana\address[INAFMI]{Osservario Astronomico di Brera, \\
	\hspace{\apex}Via Bianchi 46, 23807 Merate (LC), Italy},
	S.~Del~Sordo\address[IASFPA]{Istituto di Astrofisica Spaziale e
	Fisica Cosmica (IASF/CNR), Sezione Palermo, \\
	\hspace{\apex}Via La~Malfa 153, 90146 Palermo, Italy},
	D.~de~Martino\address[INAFNA]{Osservario Astronomico di Capodimonte, \\
	\hspace{\apex}Via Moiariello 16, 80131 Napoli, Italy},
	F.~Frontera\addressmark[IASFBO]\thanks{Also Dipartimento di
	Fisica, Universit\`a di Ferrara, Via Paradiso 11, 44100 Ferrara, Italy},
	A.~Guarnieri\addressmark[INAFBO],
	G.~Israel\address[INAFRO]{Osservario Astronomico di Monteporzio Catone,
	\\ \hspace{\apex}Via Frascati 33, 00040 Monteporzio Catone (RM), Italy},
	N.~Masetti\addressmark[IASFBO],
	E.~Palazzi\addressmark[IASFBO],
	A.~Piccioni\addressmark[INAFBO],
	A.~Santangelo\addressmark[IASFPA] and
	L.~Stella\addressmark[INAFRO]
       }
       
\begin{document}

\begin{abstract}
We present the result of three BeppoSAX observations of the X--ray binary
pulsar \s\ in its quiescent state.  The source is quite well detected up
to 200 keV ($6\sigma$ detection in the stronger observation).  No Iron
line is detected in the MECS data ($3\sigma$ upper limit on its equivalent
width of 150 eV). There is evidence of a soft excess below 2 keV. 
Pulsation is detected in all energy bands up to 10 keV, with a pulsed
fraction of $\sim 0.5$, not varying with energy.  There is a marginal
detection ($4\sigma$) of a cyclotron resonance feature (CRF) at $118 \pm
20$ keV in the PDS spectrum. During the BeppoSAX observations HDE~245770,
the optical counterpart to \s, was monitored spectroscopically and
photometrically. These observations show that the $H_\alpha$ and $H_\beta$
lines previously observed in absorption returned in emission, indicating
that the Be disk formed again.  The presence of pulsation and a CRF is a
clear indication that accretion onto the polar caps is occurring, and that
the propeller mechanism is leaking. The presence of a soft excess could be
explained either in terms of thermal emission ($kT\sim 1.2$ keV) from the
neutron star surface, as observed in other accreting X--ray pulsars, or in
terms of an overabundance of Mg in the circumstellar matter.
\vspace{1pc}
\end{abstract}

\maketitle

\section{INTRODUCTION}

The class of High Mass X--ray binaries containing a Be star (Be/HMXRBs) 
is characterized by transient X--ray outbursts, sometimes reaching
Eddington luminosities, interleaved by intervals of low luminosity ($\Lx
\le 10^{35}$ erg/s). It is possible to distinguish two types of outbursts
\cite{702}: Type~I outbursts, lasting few days and due to the increase of
the accretion rate onto the neutron star as it approaches the periastron
along the eccentric ($e\ge 0.5$) orbit;  and Type~II outbursts (also
called ``giant'' outbursts), lasting tens of days and thought to be due to
sporadic episodes of mass ejection from the Be star.  The details of the
accretion onto the neutron star are complicated by the poorly known
outflow mechanism in Be stars: because of the strong spinning of these
stars, close to 70\% of their break-up velocity \cite{2736}, circumstellar
material will accumulate in the equatorial region in the form of a
quasi-Keplerian disk.  Type~I outbursts are thought to originate when the
neutron star passes through this slow, dense disk.  Quite recently,
theoretical works \cite{2635,2458} have shown that tidal interactions can
produce truncation of the Be disk at a distance smaller than the
periastron separation, suppressing the X--ray activity, and therefore
explaining the missing of some Type~I outbursts. In this model, long term
($\sim$ years) cycles arise as stellar material accumulate in the Be disk
until tidal interactions sweep it away. The precession/warping of the Be
disk accounts for the irregularity of the Type~II outburst recurrences. 

The study of the quiescent emission in Be/HMXRBs can give important clues
on the accretion flow geometry because of the tremendous difference in
luminosity in the quiescent and outburst states (more than three orders of
magnitude). In particular, we expect that the region responsible for the
X--ray emission passes from being optically thick at high luminosity to be
optically thin in quiescence, because of the formation of a collision-less
shock above the polar caps of the neutron star during outburst states
\cite{898}.  These variations in X--ray luminosity are responsible of
dramatic variations in the pulse shape \cite{1734}, because for $\Lx >
10^{36}$ erg/s X--ray photons escape mainly in a direction perpendicular
to the magnetic field lines (fan beam emission pattern), while at lower
luminosities photons are emitted mainly in a direction parallel to the
magnetic field lines (pencil beam).  The comparison of pulse shapes at
different luminosities (quiescent vs outburst phase, but also at different
orbital phases during quiescence) can give therefore some clues on the
emission pattern geometry. 

Furthermore, the study of the quiescent state in Be/HMXRBs can give
important information on the properties of mass accretion at very low
accretion rates, and on the possible onset of a propeller state (see
Section~\ref{discussion} below), in which the accretion onto the neutron
star is centrifugally inhibited. 

In this paper we present the results of a campaign of observations
performed by BeppoSAX in order to study the quiescent emission of \s\ at
different orbital phases, and therefore at different luminosity levels. 
Simultaneous optical observations of HDE~245770 performed at the 1.5m
Cassini Telescope of the Bologna Astronomical Observatory are also
presented.

\section{\s}

\s\ is the prototype of the Be/HMXRBs. It was discovered during a Type~II
outburst in 1975 \cite{747,748}, and since then three other giant
outbursts and about a dozen of normal outbursts have been observed
\cite{377}.  X--ray pulsations at $\sim$104~s have been detected both
during outburst \cite{752} and in quiescence \cite{48}.  Because of the
long database of observations it was possible to resolve the binary system
\cite{1397}, that exhibits an orbital period of 111 days, and an
eccentricity of 0.47. The optical counterpart to \s\ is HDE~245770
\cite{2737}, of spectral type O9.7IIIe--B0Ve \cite{2738}.  The Be star
presents $UBV$ photoelectric variability correlated to the X--ray behavior
of \s\ \cite{2177}: in particular the $V$ luminosity reached its maximum
in 1975 ($V\sim 8.8$ \cite{2690}), in correspondence of a giant X--ray
outburst. The minimum was observed in 1998 ($V\sim 9.1$ \cite{2690}), when
the optical spectrum showed a dramatic change, with the emission lines
slowly changing from emission to absorption.  This phenomenon is
interpreted in terms of Be disk loss \cite{2042}.

\section{OBSERVATIONS AND RESULTS}

\subsection{X--ray observations}

\s\ was observed three times at different orbital phases by the Narrow
Field Instruments (NFIs) aboard the Italian-Dutch satellite BeppoSAX
\cite{1530}: LECS (0.5--10 keV \cite{1531}), MECS (1.8--10 keV
\cite{1532}), and PDS (15--200 keV \cite{1386}).  During the whole
campaign the HPGSPC \cite{1533} was switched off.  The log of the
observations, together with the observed NFIs count rates are reported in
Table~\ref{log}. The first two observations, in which the neutron star
approaches the periastron, show the same intensity, while the third
observation, after the periastron passage, has an almost double count rate
in all the NFI energy bands. 

% Lengths to be used in the table
\newlength{\lenOP}   \settowidth{\lenOP}{00000}
\newlength{\lenDT}   \settowidth{\lenDT}{00/00/00 00:00:00M}
\newlength{\lenthr}   \settowidth{\lenthr}{61811M}

\begin{table*}
\caption{Log of the observations of \s\ performed by BeppoSAX}
\label{log}
\begin{tabular}{lll@{\hspace{-0.4em}}l@{\hspace{2em}}cccc}
\noalign{\medskip}\hline\noalign{\medskip}
& \mcc{OP} & \mcc{Start Time}  & \mcc{Length} & \multicolumn{3}{c}{\Iave\ (10$^{-2}$ C/s)} & Orbital \\ \cline{5-7}
& \mcc{\#} & \mcc{(UT)}        & \mcc{(sec)}  & \mcc{LECS} & \mcc{MECS} & \mcc{PDS}        & Phase$^a$ \\
& 	   &                   &              & \mcc{(0.5--2)}  & \mcc{(2--10)}  & \mcc{(15--200)} & \\
\noalign{\medskip}\hline\noalign{\medskip}
OBS1 &
\begin{minipage}{\lenOP}09736\\ 09738\end{minipage} &
\begin{minipage}{\lenDT}04/09/00 05:14:05\\ 05/09/00 00:41:56\end{minipage} &
\hfill \begin{minipage}{\lenthr}
$\!\left. \begin{array}{r} ~~68865\\ 50020\end{array}\!\!\! \right\}$
\end{minipage}
		& $0.56\pm 0.06$ & $3.2\pm 0.1$ & $20\pm 6$ & 0.606--0.618 \\
\noalign{\medskip}\noalign{\medskip}
OBS2 &
\begin{minipage}{\lenOP}09923\\ 09924\end{minipage} &
\begin{minipage}{\lenDT}05/10/00 00:42:46\\ 06/10/00 04:29:26\end{minipage} &
\hfill \begin{minipage}{\lenthr}
$\!\left. \begin{array}{r} 100000\\ 94584\end{array}\!\!\! \right\}$
\end{minipage}
                & $0.56\pm 0.05$ & $3.1\pm 0.1$ & $11\pm 5$ & 0.883--0.903 \\
\noalign{\medskip}\noalign{\medskip}
OBS3 &
\begin{minipage}{\lenOP}10835\\ 10836\\ 10837\end{minipage} &
\begin{minipage}{\lenDT}05/03/01 22:52:38\\ 07/03/01 02:39:28\\ 08/03/01 06:25:58\end{minipage} &
\hfill \begin{minipage}{\lenthr}
$\!\left. \begin{array}{r} 100000\\ 100000\\ 40415\end{array}\!\!\! \right\}$
\end{minipage}
		& $1.20\pm 0.05$ & $7.8\pm 0.1$ & $24\pm 4$ & 0.251--0.276 \\
\noalign{\medskip}\hline\noalign{\medskip}
\multicolumn{7}{l}{$^a$ Ephemeris from Finger (private communication): $T_N = 45948.8(16)+N\times 111.07(7)$}
\end{tabular}
\end{table*}

\subsubsection{Timing analysis}

\begin{figure*}
\centerline{\includegraphics[angle=270,width=0.9\textwidth]{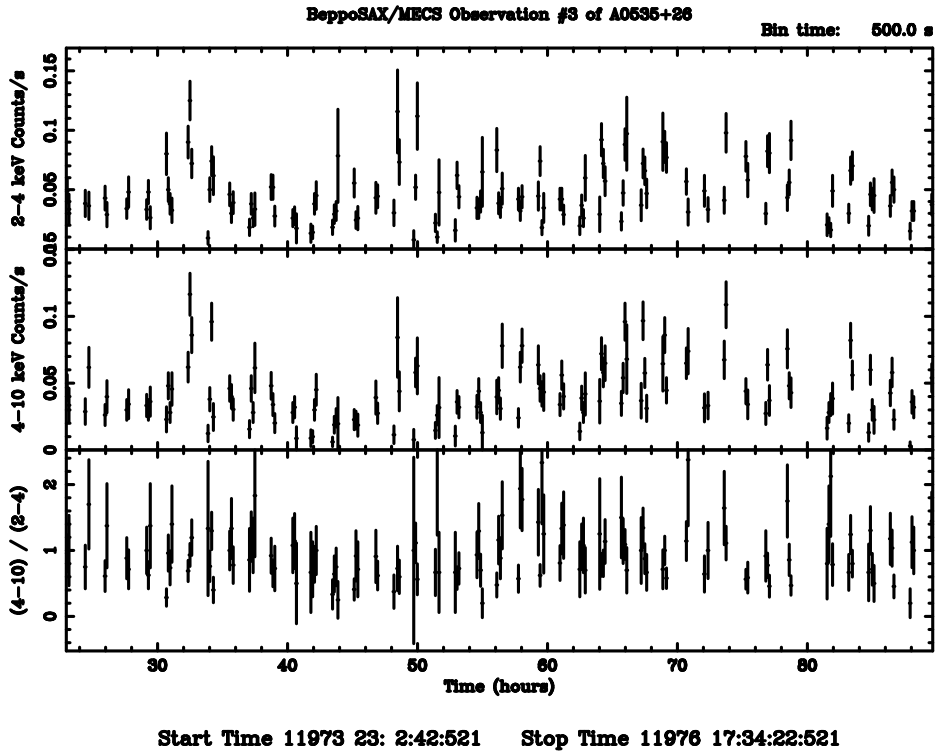}}
\vspace{-2\baselineskip}
\caption[]{500~s background subtracted MECS light curves of the third
BeppoSAX observation in two energy bands (first two panels) and their
ratio (lower panel).  There is no appreciable spectral evolution during the
whole observation}
\label{lc}
\end{figure*}

During each BeppoSAX observation the source was quite stable, with time
variability on time scales of the order of hundreds of seconds and
intensity variability within a factor two (see Fig.~\ref{lc} with the OBS3
light curve as observed in two MECS energy bands. Gaps in the data are due
to passages in the South Atlantic Geomagnetic Anomaly, when the NFIs are
switched off). The $(4-10)$/$(2-4)$ keV hardness ratio does not show any
variation during the observations.

\begin{figure}
\centerline{\includegraphics[angle=270,width=\columnwidth]{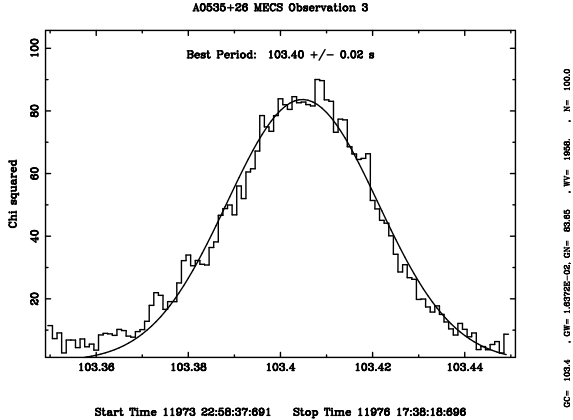}}
\vspace{-2\baselineskip}
\caption[]{Epoch folding search result on the third \s\ observation. A
Gaussian fit to the curve was used to compute the error on the measured
spin period}
\label{efs}
\end{figure}

After correcting photon arrival times to the solar system barycenter, MECS
data were searched for periodic signals by performing an epoch folding
search and the result is shown in Fig.~\ref{efs} for OBS3.  Data were not
corrected for the \s\ orbital motion because of the short duration of the
observation with respect to the orbital period of 111 days.  Our result of
$103.40\pm 0.02$~s is in agreement with the other quiescent pulse period
measurements \cite{377,2084}, confirming that the source is accreting
material and channeling it onto the neutron star polar caps even when in
quiescence.

\begin{figure*}
\hspace{0.5cm}{\includegraphics[height=\textwidth]{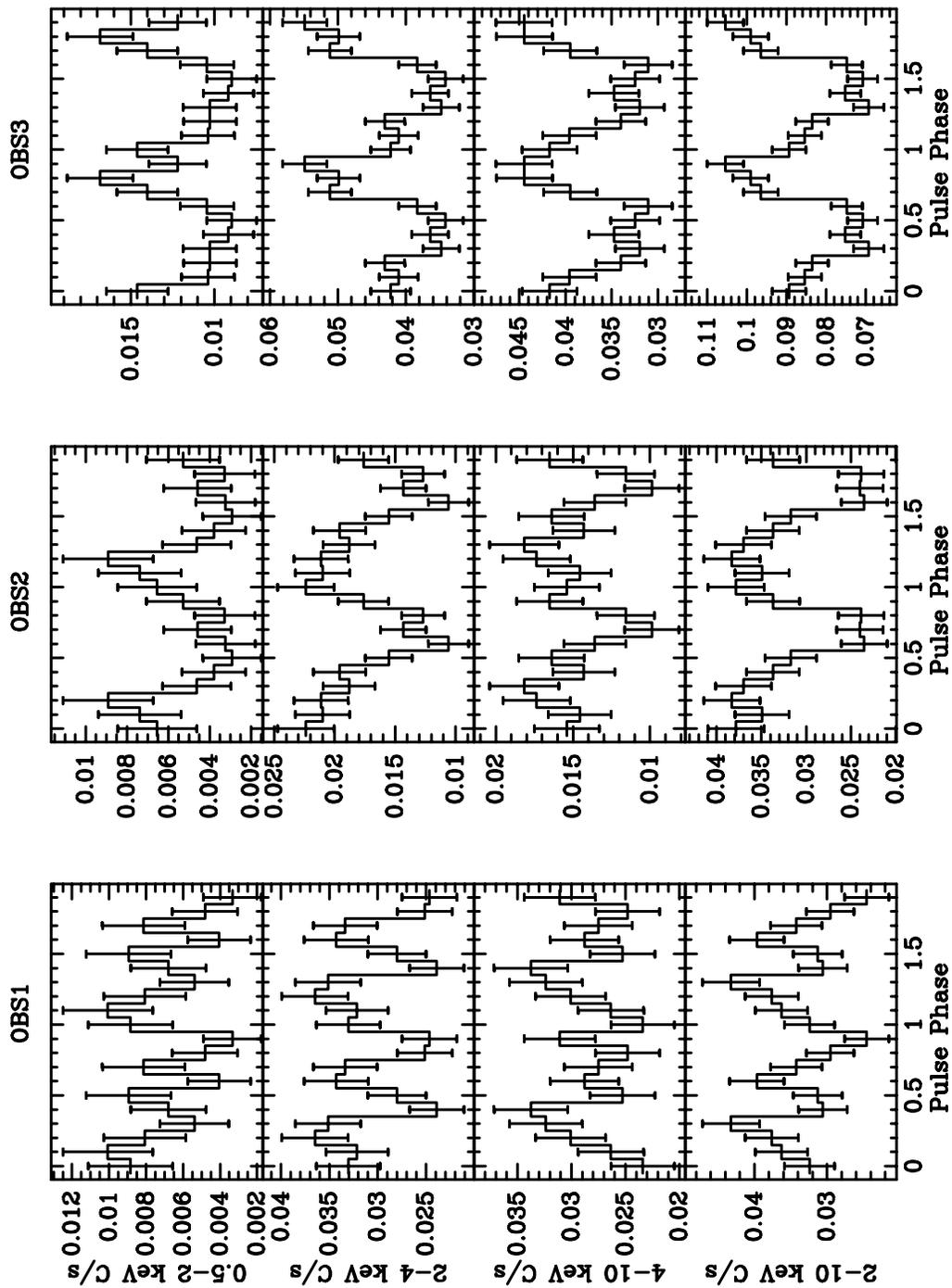}}
\vspace{3\baselineskip}
\caption[]{Background subtracted LECS and MECS pulse profiles for all the
three \s\ observations in four energy bands: 0.5--2 keV (LECS; first row),
2--4 and 4--10 keV (MECS; second and third rows).  In the last row it is
shown the 2--10 keV pulse profile. All the profiles are phased at MJD
11791}
\label{pulseprofiles}
\end{figure*}

\begin{figure}
\centering{\includegraphics[width=\columnwidth]{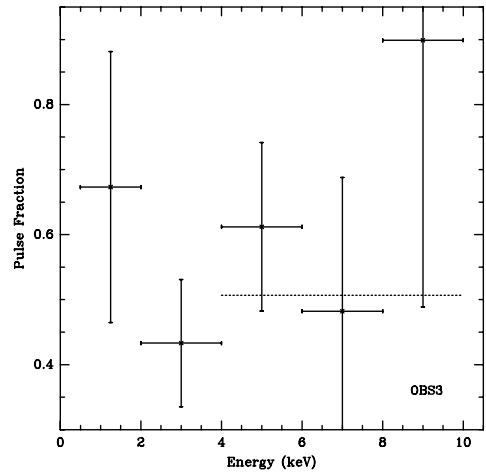}}
\vspace{-5\baselineskip}
\caption[]{Pulse fraction as a function of energy for OBS3. The dotted
line represents the pulse fraction measured on the whole 4--10 keV energy
range}
\label{pulsefraction}
\end{figure}

In Fig.~\ref{pulseprofiles} we show the LECS and MECS pulse profiles as a
function of energy for all the three observations (we were not able to
extract pulse profiles from the PDS instrument, because of the very low
statistics). In the first observation the source displays a double peak
profile, typical of the high energy band \cite{302,1734}. In OBS2 the
pulse is clearly single peaked, of the same kind observed by RXTE at
orbital phase 0.64 \cite{2084}, while in the third it is structured, with
the presence of a pronounced first peak and a smaller second one.  The
pulse fraction within the uncertainties, does not depend on energy, as it
is shown in Fig.~\ref{pulsefraction} for OBS3.  Our result is in agreement
with the pulse fraction of 0.53 measured by \cite{2084}.

\begin{table*}
\caption{\s\ LECS+MECS spectral fits$^a$}
\label{fits}
\begin{tabular}{lccc}
\noalign{\medskip}\hline\noalign{\medskip}
 & OBS1 & OBS2 & OBS3 \\
\noalign{\medskip}\hline\noalign{\medskip}
{\bf Power Law} \\
N$_H$ (10$^{22}$ cm$^{-2}$) & $1.10\pm 0.19$ & $1.03\pm 0.13$ & $1.33\pm
0.09$ \\
$\alpha$                    & $1.98\pm 0.11$ & $1.92\pm 0.09$ & $1.85\pm
0.05$ \\
$\chi^2_\nu$ (dof)          & 0.90(68)       & 1.17(98)       &
1.33(208) \\
\noalign{\medskip}\hline\noalign{\medskip}
{\bf Bremsstrahlung} \\
N$_H$ (10$^{22}$ cm$^{-2}$) & $0.75\pm 0.14$ & $0.75\pm 0.10$ & $1.02\pm
0.07$ \\
$kT$ (keV)                  & $7.4\pm 1.2$ & $7.7\pm 1.1$ & $9.4\pm 0.8$
\\
$\chi^2_\nu$ (dof)          & 0.90(68)       & 1.07(98)       &
1.17(208) \\
\noalign{\medskip}\hline\noalign{\medskip}
{\bf Black Body} \\
$kT$ (keV)                  & $1.09\pm 0.03$ & $1.14\pm 0.02$ & $1.26\pm
0.01$ \\
$R$ (Km)                    & $0.088\pm 0.004$ & $0.080\pm 0.003$ &
$0.106\pm 0.002$ \\
$\chi^2_\nu$ (dof)          & 1.37(69)       & 1.23(99)       &
1.30(209) \\
\noalign{\medskip}\hline\noalign{\medskip}
\Lx\ (0.5--2 keV)$^b$       & 0.6            & 0.5            & 1.2 \\
\Lx\ (2--10 keV)$^b$        & 1.5            & 1.5            & 4.4 \\
\noalign{\medskip}\hline\noalign{\medskip}
\multicolumn{4}{l}{$^a$ Uncertainties at 90\% confidence level for a
 single parameter} \\
\multicolumn{4}{l}{$^b$ Un-absorbed luminosities in $10^{33}$ erg/s
 assuming a distance $d = 2$ kpc \cite{1682}}
\end{tabular}
\end{table*}

\begin{figure*}
\centerline{\includegraphics[angle=270,width=0.65\textwidth]{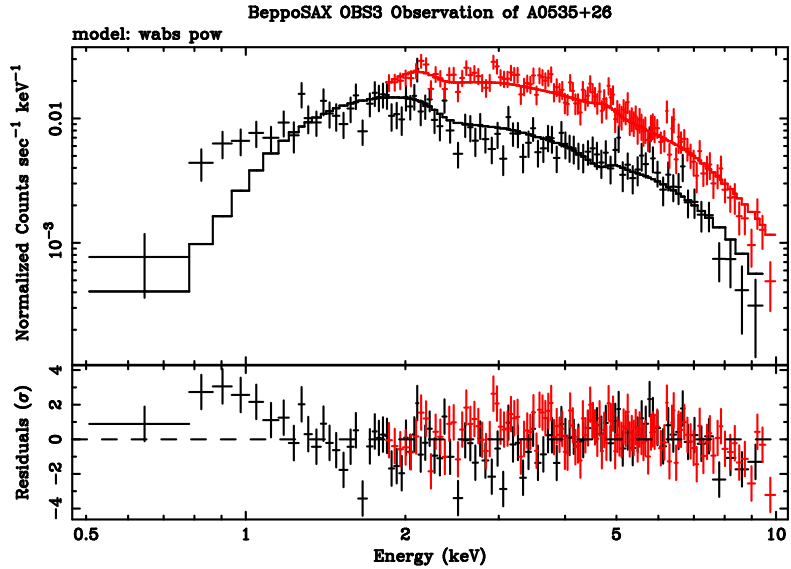}}
\vspace{-2\baselineskip}
\caption[]{BeppoSAX \s\ count rate spectrum ({\em plus signs}\/; LECS in
black and MECS in gray) and power law best-fit continuum ({\em
histogram}\/), together with the fit residuals.  Fitting parameters are
listed in Table~\ref{fits}}
\label{mecs-fit}
\end{figure*}

\subsubsection{Spectral analysis}

The LECS and MECS energy spectra for all the observations were extracted
using standard procedures, as outlined in \cite{1827}.  The combined
LECS/MECS spectra are well fit both by an absorbed power law and a thermal
bremsstrahlung. Also a blackbody fit is statistically acceptable, but the
resulting emitting region is unphysically small (see \cite{2004} on this
topic), and no absorption is required.  These results are in agreement
with \cite{2084}, and are listed in Table~\ref{fits}.  In
Fig.~\ref{mecs-fit} we show the power law fit for the third observation,
in which a soft excess below 2~keV is quite evident. This excess can be
modeled both with the addition of a black body component ($kT = 1.2$ keV;
$\chi^2$/dof = 213.5/206), or by an over-abundance of Mg in the
circumstellar matter ($14\pm 3$ times the Solar abundance; $\chi^2$/dof =
232.5/208).  Our data do not show the presence of an iron $K$-shell line
in 6.4--6.9 keV: the 3$\sigma$ upper limit on its equivalent width is
150~eV. 

For extraction of the PDS net spectra we used the standard procedure
\cite{1586} for all the observations but the third where one of the offset
fields used for the determination of the background was contaminated by
the Crab nebula, and therefore could not be used in the pipeline.  In all
the observations \s\ is clearly detected in the PDS (see Table~\ref{log})
up to 200 keV.  To our knowledge, this is the first detection up to these
energies of the quiescent state of \s.  Because of the low statistics only
a power law model could reasonably be applied for fitting the PDS spectra. 
The best fitting results are listed in Table~\ref{fit-pds}.  These results
indicate a hardening of the spectrum for E$>$15~keV consistent with
$\alpha=1.1-1.4$ observed by HEXE (15--200 keV) on-board MIR during an
outburst \cite{1734}. 

\begin{figure}
\includegraphics[angle=270,width=1.1\columnwidth]{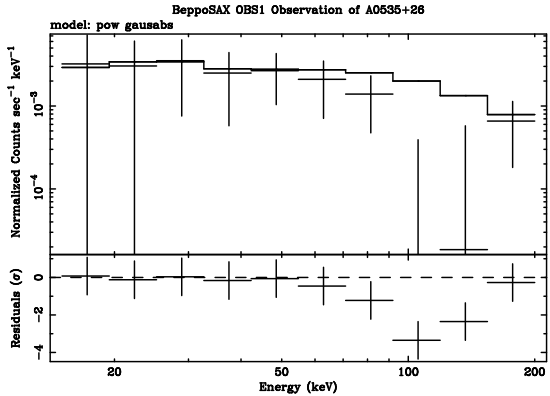}
\vspace{-3\baselineskip}
\caption[]{PDS 15--200 keV count rate spectrum for OBS1. Top panel:  count
rate spectrum ({\em plus signs\/}) and best fit continuum model ({\em
histogram\/}). Bottom panel: residuals to the power law plus Gaussian in
absorption model.  The CRF normalization was put to zero, in order to show
the feature significance}
\label{fig:fit-PDS}
\end{figure}

\begin{table}
\caption[]{A0535+26 PDS power law spectral fits}
\label{fit-pds}
\begin{tabular}{lccc}
\noalign{\medskip}\hline\noalign{\medskip}
 & OBS1 & OBS2 & OBS3 \\
\noalign{\medskip}\hline\noalign{\medskip}
$\alpha$                    & $1.0\pm 0.5$ & $2.1\pm 0.8$ & $1.0\pm 0.2$ \\
$\chi^2_\nu$ (dof)          & 0.55(8)      & 0.86(8)      & 0.86(8) \\
\noalign{\medskip}\hline\noalign{\medskip}
\multicolumn{4}{l}{\bf CRF best fit parameters} \\
$\alpha$                    & $0.5^{+0.4}_{-0.8}$    \\
$E_{\rm CRF}$ (keV)         & $118^{+21}_{-19}$      \\
$\sigma$ (keV)              & $20^{+20}_{-19}$       \\
E.W.\ (keV)                 & $81^{+50}_{-48}$       \\
\noalign{\medskip}\hline\noalign{\medskip}
\end{tabular}
\end{table}

The PDS spectra for OBS2 and OBS3 do not show any deviation from the power
law fit, while the OBS1 spectrum presents structured residuals in the
range 100--200 keV (see the slightly worse $\chi^2_\nu$ for OBS1 in
Table~\ref{fit-pds}).  Because of a previous observation by OSSE
\cite{375} of a cyclotron resonance feature (CRF) at 110 keV (observed
during a giant outburst), we tried to model the feature with a Gaussian in
absorption \cite{1172}, as it has been done with other CRFs observed by
BeppoSAX. We prefer this kind of model with respect to the Lorenzian model
\cite{1547} because the latter has the disadvantage of modifying the
continuum (see discussion in \cite{1502}).  In Fig.~\ref{fig:fit-PDS} we
show the power law plus Gaussian in absorption fit for OBS1, where in the
residuals panel we put the CRF normalization to zero in order to show its
significance.  The CRF best fit parameters are listed in
Table~\ref{fit-pds}, in agreement with the OSSE results.

\begin{figure}
\qquad\centerline{\includegraphics[width=\columnwidth]{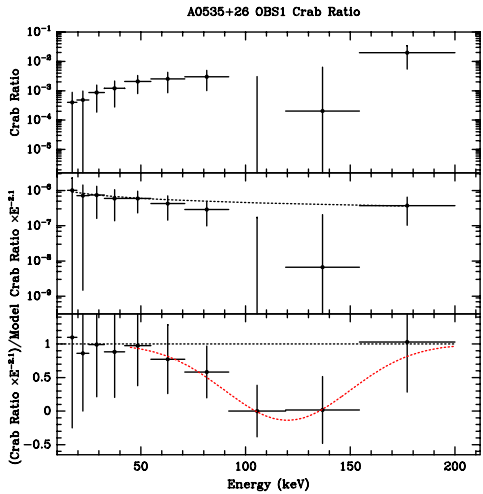}}
\vspace{-4.5\baselineskip}
\caption[]{Normalized Crab ratio for the first PDS observation of \s. 
First panel: ratio between the \s\ and Crab count rate spectra.  Middle
panel: Crab ratio multiplied by the functional form of the Crab spectrum,
a E$^{-2.1}$ power law. Bottom panel: The result from the previous
operation divided by the functional describing the continuum (as listed in
Table~\ref{fit-pds})}
\label{crab-ratio}
\end{figure}

In order to check whether the observed spectral feature was real or due to
an instrumental effect, we performed a normalized Crab ratio \cite{1622}
on the PDS count rate spectrum of the first observation.  This technique
has been successfully applied for the detection of CRFs in other X--ray
pulsars \cite{1502} and consists of the ratio between the source count
rate spectrum and the count rate spectrum of the Crab Nebula. As this
second spectrum is known, with great accuracy, to be free of features and
to be modeled at first order with a power law in a very broad energy
range, this ratio is quite well suited to enhance the presence of features
in the spectrum.  Furthermore the ratio is in first approximation
independent from the calibration of the instrument.  In order to enhance
the deviations from the continuum we multiply the ratio by a E$^{-2.1}$
power law, that is the functional form of the Crab spectrum, and we divide
by the functional describing the continuum shape of the source (from this
the name normalized Crab ratio). The procedure is described in
Fig.~\ref{crab-ratio} where we plot the result of each different step used
to obtain the final result in the case of \s.

\subsection{Optical observations}

\begin{figure*}
\centerline{\includegraphics[angle=270,width=0.8\textwidth]{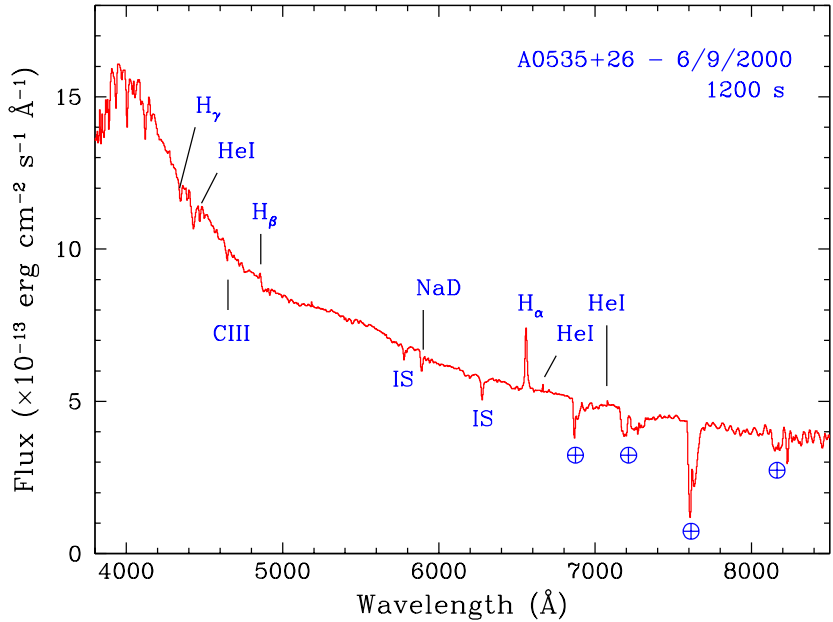}}
\vspace{-3\baselineskip}
\caption[]{Optical spectrum of HDE~245770 taken at the 1.5m Cassini
telescope of the Bologna Astronomical Observatory. The star returned to
its {\em normal} Be spectrum}
\label{opt}
\end{figure*}

The optical counterpart to \s, the Be star HDE~245770, was monitored at
the 1.5m Cassini telescope of the Bologna Astronomical Observatory almost
continuously since its discovery in 1975 \cite{2691}.  During the BeppoSAX
campaign several medium-resolution optical spectra, with dispersion
ranging from 1.8 to 4 \AA/pixel and covering the 4000-8500 \AA~range, were
acquired simultaneously with the X--ray pointings by using the BFOSC
instrument. One of these spectra is shown in Fig.~\ref{opt}. In all of
them it is clearly seen that H$_\alpha$ and H$_\beta$ are in emission,
that is the Be star spectrum returned to its {\em normal} Be spectrum. 
This is in agreement with the rebuilding of the Be circumstellar disk
around HDE~245770.

At the time of these observations, optical photometry data collected at
the nearby Loiano 60 cm telescope shows that the star was at magnitude
$V\sim$ 9.0 \cite{giovannelli03}. 

\section{DISCUSSION \label{discussion}}

Recent RXTE observations \cite{2084} of the quiescent emission from \s\ at
$\Lx\simeq 4\times 10^{33}$ erg/s (2--10 keV; two orders of magnitude
lower than that observed in past quiescent states \cite{48}) pose serious
problems for the standard accretion theory (see, e.g.\ \cite{870}).
Indeed, at this luminosity level the corotation radius (the distance from
the neutron star at which the centrifugal force equals the gravitational
pull) is smaller than the magnetospheric radius (the distance at which the
magnetic pressure equals the ram pressure), and therefore matter cannot
enter the magnetosphere and being accreted because of centrifugal forces
(this is the so called propeller state \cite{873}).  But the detection of
X--ray pulsations clearly contradicts that the source is in a centrifugal
inhibited state.  Optical observations before and during the RXTE
observations also showed that the $JHK$ infrared excess and IR/optical
lines emission were missing \cite{2042}, a clear indication of a Be disk
loss episode in HDE~245770. 

In order to accommodate this discrepancy Ikhsanov \cite{2301} proposed that
the mechanism of plasma penetration into the magnetosphere is not through
standard interchange instabilities, but through magnetic reconnection of
the field lines.  In this way, both the presence of pulsation and the
small polar cap radius derived from the RXTE spectral fits are explained.

Our X--ray observations confirm the low luminosity state of the source,
and therefore its centrifugally inhibited state.  The LECS observations
were able to clearly detect a soft excess that could be explained both in
terms of an additional thermal component, as already observed in other
accreting X--ray binary pulsars, like Her X--1 \cite{1583}, or
overabundance of Mg ($\sim10$ times the Solar abundance) in the proximity
of the source.  The almost absence of spectral variability during the
orbit seems to point to the absence of any (temporary) accretion disk, as
observed during a giant outburst phase from the detection of spin and QPO
frequency variations \cite{1397}.

%Our attempts to model the soft excess
%with thermal emission from an accretion disk ({\tt diskbb} in XSPEC)
%give an unphysical value of the inner disk radius $R_{\rm
%in}^2\cos\theta = 6\times 10^{-4}$ Km$^2$, where $\theta$ is the angle
%subtended by the disk.

The main difference between BeppoSAX and RXTE observations is that the Be
disk around HDE~245770 returned in place.  We would have expected a higher
X--ray luminosity with respect to the disk-less observations, due to the
greater amount of circumstellar matter that can be accreted.  Therefore
the presence of the Be disk seems not to alter the accretion rate onto the
neutron star.

The high energy observations indicate a spectral hardening, which is in
disagreement with the X--ray spectra observed during the \s\ active phase,
characterized by a spectral softening above 25~keV \cite{375}.  This is
further evidence that the kind of accretion in the quiescent and the
active phase seems different.  In the first BeppoSAX observation there is
a marginal (4$\sigma$) evidence of a CRF at 110~keV, but not of a 55~keV
feature (reported by \cite{1346}). This is a confirmation of the OSSE
observation of the CRF detected during the 1994 giant outburst. The same
CRF energy observed in two completely different luminosity states
constrains the physical conditions in the region in which the CRF
originates to not depend on the accretion rate, and this poses serious
limits on the actual emission models. 

\medskip

\noindent {\em Acknowledgments.} \quad We do honor to our late colleague
Daniele Dal~Fiume, Principal Investigator of the \s\ BeppoSAX
observations, for his invaluable contribution, among others, in the field
of accreting X--ray pulsars. This research was partially funded by Agenzia
Spaziale Italiana.

\end{document}